\documentclass[fleqn,12pt]{wlscirep}
\usepackage{diagbox}

\newcommand{\fhi}{\overline{x}_{\mathrm{H}\ensuremath{\textsc{i}}}}

\newcommand{\hi}{H\ensuremath{\,\textsc{i}}}

\newcommand{\mgii}{Mg\ensuremath{\,\textsc{ii}}}
\newcommand{\cii}{[C\ensuremath{\,\textsc{ii}}]}

\newcommand{\civ}{C\ensuremath{\,\textsc{iv}}}

 \newcommand{\kms}{{\rm km\,s}\ensuremath{^{-1}}}

\newcommand{\aap}{Astron. Astrophys.}               % Astronomy and 
\newcommand\apjs{Astrophys. J. Suppl. Ser.}%
\newcommand\apjl{Astrophys. J}%

\title{An 800-million-solar-mass black hole in a significantly neutral Universe at redshift 7.5}

\author[1,*]{Eduardo Ba\~nados}
\author[2]{Bram P. Venemans}
\author[2]{Chiara Mazzucchelli}
\author[2]{Emanuele~P.~Farina}
\author[2]{Fabian Walter}
\author[3,4]{Feige Wang}
\author[2,5]{Roberto Decarli}
\author[6]{Daniel Stern}
\author[7]{Xiaohui Fan}
\author[8]{Frederick B. Davies}
\author[8]{Joseph F. Hennawi}
\author[9]{Robert A. Simcoe}
\author[9,10]{Monica L. Turner}
\author[2]{Hans-Walter Rix}
\author[3,4]{Jinyi~Yang}
\author[1]{Daniel D. Kelson}
\author[1]{Gwen C. Rudie}
\author[11]{Jan Martin Winters}

\affil[1]{The Observatories of the Carnegie Institution for Science, 813 Santa Barbara Street., Pasadena, CA 91101, USA}
\affil[2]{Max Planck Institut f\"ur Astronomie, K\"onigstuhl 17, D-69117, Heidelberg, Germany}
\affil[3]{Department of Astronomy, School of Physics, Peking University, Beijing 100871, China}
\affil[4]{Kavli Institute for Astronomy and Astrophysics, Peking University, Beijing 100871, China}
\affil[5]{INAF -- Osservatorio Astronomico di Bologna, via Gobetti 93/3, 40129, Bologna, Italy}
\affil[6]{Jet Propulsion Laboratory, California Institute of Technology, 4800 Oak Grove Drive, Pasadena, CA 91109, USA}
\affil[7]{Steward Observatory, The University of Arizona, 933 North Cherry Avenue, Tucson, AZ 85721--0065, USA}
\affil[8]{Department of Physics, Broida Hall, University of California, Santa Barbara, CA 93106--9530, USA}
\affil[9]{MIT-Kavli Center for Astrophysics and Space Research, 77 Massachusetts Avenue, Cambridge, MA, 02139, USA}
\affil[10]{Las Cumbres Observatory, 6740 Cortona Dr, Goleta, CA 93117, USA}
\affil[11]{Institut de Radioastronomie Millimétrique (IRAM), 300 rue de la Piscine, 38406 Saint Martin d’Hères, France}

\affil[*]{ebanados@carnegiescience.edu}

\begin{abstract}
\textbf{Quasars are the most luminous non-transient objects known and as a result they enable studies of the Universe at the earliest cosmic epochs. 
Despite extensive efforts, however, the quasar ULAS~J1120+0641 at $z=7.09$ has remained the only one known at $z>7$ for more than half a decade\cite{mortlock2011}. 
Here we report observations of the quasar ULAS~J134208.10+092838.61 (hereafter J1342+0928) at  redshift  $z=7.54$. 
 This quasar has a bolometric luminosity of $4\times 10^{13}$ times the luminosity of the Sun and a black hole mass of $8 \times 10^8$ solar masses. The existence of this supermassive black hole when the Universe was only 690 million years old---just five percent of its current age---reinforces models of early black-hole growth that allow black holes with initial masses of more than about $10^4$ solar masses\cite{latif2013b,alexander2014}
  or episodic hyper-Eddington accretion\cite{pacucci2015,inayoshi2016b}. 
We see strong evidence of absorption of the spectrum of the quasar redwards of the Lyman $\alpha$ emission line (the Gunn-Peterson damping wing), as would be expected if a significant amount (more than 10 per cent) of the hydrogen in the intergalactic medium surrounding J1342+0928 is neutral. We derive a significant fraction of neutral hydrogen, although the exact fraction depends on the modelling. However, even in our most conservative analysis we find a fraction of more than 0.33 (0.11) at 68 per cent (95 per cent) probability, indicating that we are probing well within the reionization epoch of the Universe. 
}

\end{abstract}

\begin{document}

\flushbottom
\maketitle
\thispagestyle{empty}

We detected the quasar J$1342+0928$ as part of an on-going effort to find quasars at $z>7$ by mining three large-area surveys: the \textit{Wide-field Infrared Survey Explorer}\cite{wright2010} (ALLWISE), the United Kingdom Infrared Telescope Infrared Deep Sky Survey (UKIDSS) Large Area Survey\cite{lawrence2007}, and the DECam Legacy Survey (DECaLS; http://legacysurvey.org/decamls). 
At redshifts of more than about $7$, residual neutral hydrogen in the intergalactic medium (IGM) absorbs virtually all flux bluewards of the Lyman $\alpha$ (Ly$\alpha$) emission line, which is redshifted to observed wavelengths of greater than about one micrometre, meaning that quasars at these redshifts are not detectable in the optical bands. 
To identify quasars at redshifts greater than 7, we therefore required a detection in both UKIDSS $J$ and \textit{WISE} $W1$ bands with a signal-to-noise ratio greater than 5 and no source in the DECaLS DR3 catalogue within $3^{\prime\prime}$. We also required a non-detection in the DECaLS $z_{\rm DE}$-band image, as indicated by a flux of $z_{\rm{DE},3\sigma} - J> 2$ determined by forced photometry (here $z_{\rm{DE},3\sigma}$ is the $3\sigma$ lower limit for the magnitude of the quasar in the DECaLS  $z_{\rm DE}$-band image). Finally, we required a flat spectral energy distribution, which eliminates a large fraction of the most common contaminants of searches for quasars at redshifts of more than about  $7$: low-mass brown dwarfs in our Galaxy\cite{banados2016}. The survey photometry used to identify J$1342+0928$ is listed in Extended Data Table 1. 

\begin{figure}[h!]
\centering
\includegraphics[]{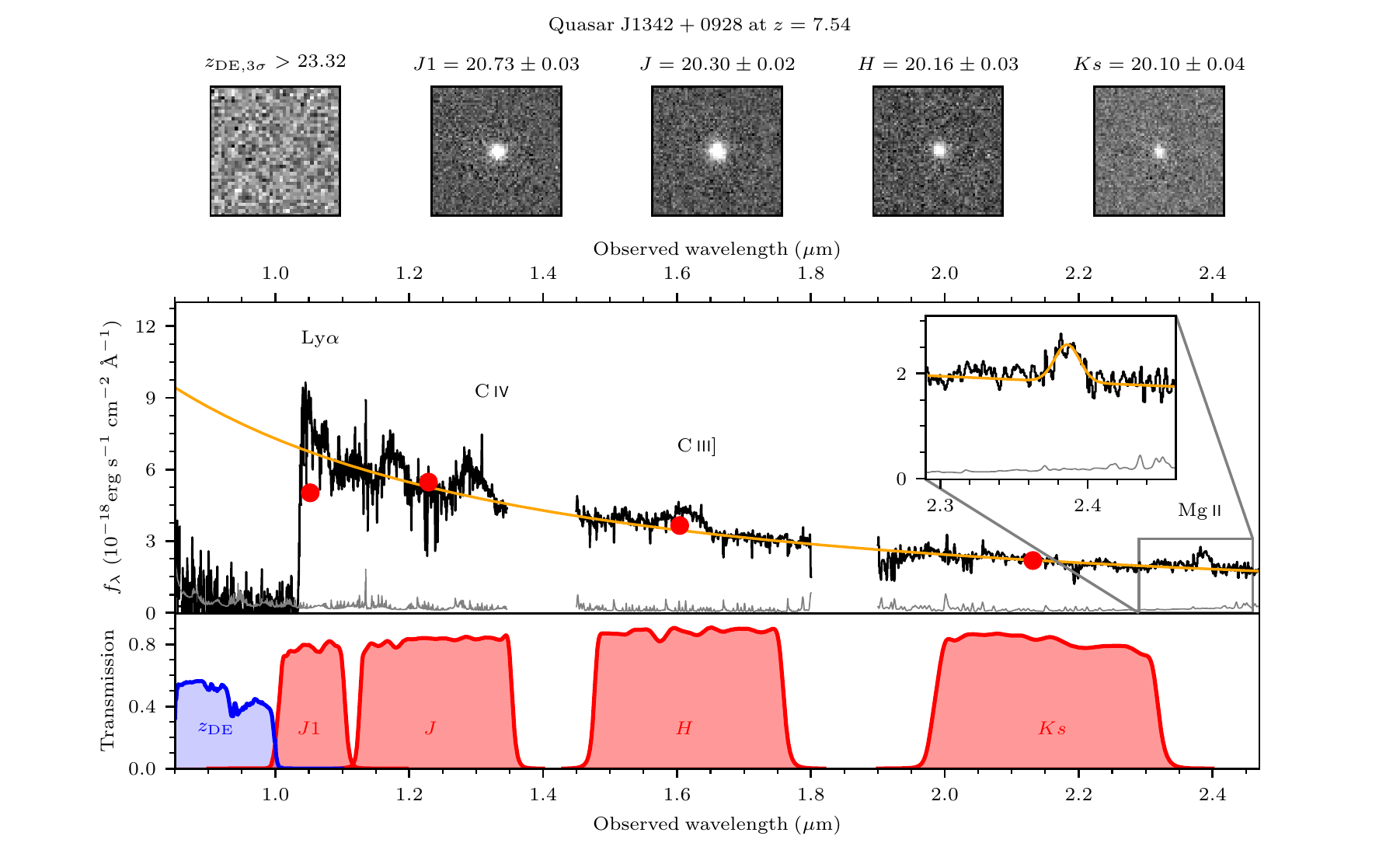}
\caption{\small
\textbf{Photometry and combined Magellan/FIRE and Gemini/GNIRS near-infrared spectrum of the quasar J1342+0928 at $z=7.54$.} 
The FIRE data were collected on 11--12 March 2017 for a total integration time of 3.5 h. We used the $0.6''$ slit in the echellete mode, yielding a spectral resolution of around $ 6,000$ over the range $0.8-2.3\, \mu$m. The GNIRS spectrum was obtained on 31 March 2017 and 3 April 2017 with a total exposure time of 4.7 h. We used the $0.675''$ slit in the cross-dispersion mode, yielding a spectral resolution of around $ 1,800$ over the range $0.8-2.5\,\mu$m. 
The spectral flux density ($f_\lambda$. black line) is shown at the GNIRS resolution, binned by a factor of two. 
The 1$\sigma$ error is shown in grey and the orange line represents the best-fitting power-law continuum emission with $f_\lambda \propto \lambda^{-1.58\pm 0.02}$.  Regions with low sky transparency between the $J-H$ and $H-Ks$ bands are not shown. The red circles show the follow-up  photometry obtained with the Magellan/Fourstar infrared camera. The inset shows a Gaussian fit to the \mgii\ line, from which we derive a black-hole mass of $7.8 \times 10^8\, M_{\odot}$. 
The bottom panel shows the transmission of the Fourstar $J1$, $J$, $H$, $Ks$ filters (red), and the DECam $z_{\rm DE}$ filter (blue); the top panel shows $10'' \times 10''$ images of the quasar in the same filters, with their respective AB magnitudes. The quasar is not detected in the $z_{\rm DE}$ image and its $3\sigma$-limiting magnitude ($z_{\rm{DE},3\sigma}$) is reported.
}
\label{fig:pisco}
\end{figure}

We confirmed J$1342+0928$ as a quasar with a 10 min spectrum with the Folded-port InfraRed Echellette (FIRE) spectrograph in prism mode at the Magellan 6.5\,m Baade telescope at Las Campanas Observatory on 9 March 2017.  To analyse the emission line properties in greater detail, we obtained deeper and higher-resolution spectra with FIRE, the Large Binocular Telescope (LBT) Utility Camera in the Infrared (LUCI) spectrograph at the Large Binocular Telescope,  and the Gemini Near-Infrared Spectrograph (GNIRS) at the Gemini North telescope. The LUCI spectrum provided the first detection of the \mgii\ emission line at $\sim$$2.4\,\mu$m but it was superseded by the higher-signal-to-noise ratio and larger wavelength coverage of the GNIRS spectrum. We also obtained deep follow-up photometry with the Magellan/Fourstar infrared camera on 19 March 2017. These data were used to bring the spectra to an absolute flux scale, compensating for slit-losses. The combined spectrum and follow-up photometry of J$1342+0928$ are shown in Fig. \ref{fig:pisco}.

The systemic redshift of this quasar is $z=7.5413 \pm 0.0007$ (all error bars reported here correspond to $1\sigma$ or the central 68\% interval of the distribution), measured using IRAM/NOEMA observations of the \cii\ 158\,$\mu$m emission line from its host galaxy\cite{venemans2017b}. The redshift measured from a Gaussian fit to the \mgii\ line (see Fig. \ref{fig:pisco}) is $7.527 \pm 0.004$, that is, blueshifted by $500\pm140\,$\kms\ with respect to the systemic redshift. This is consistent with the velocity offsets observed\cite{venemans2016} in other quasars at $z>6$. 
Adopting a cosmology\cite{planck2016xiii} with a current value of the Hubble parameter of $H_0 = 67.7 \,\mbox{km\,s}^{-1}$\,Mpc$^{-1}$, and cosmological density parameters $\Omega_M = 0.307$ and $\Omega_\Lambda = 0.693$, this quasar is situated at a cosmic age of just 690 Myr after the Big Bang---that is, when the Universe was about 10\% younger than at the redshift of the previous most distant quasar known\cite{mortlock2011}---at times when conditions in the Universe were changing rapidly\cite{planck2016XLVII}.

The mass of the central black hole of the quasar can be estimated using the quasar luminosity and the full-width at half maximum of its \mgii\ line, under the assumption that local scaling relations\cite{vestergaard2009} are still valid at high luminosity and high redshift\cite{mejia-restrepo2017}. 
The apparent ultraviolet magnitude measured at rest-frame $1,450$\,\AA\, from the quasar spectrum is $m_{1,450}=20.34\pm 0.04$, which translates to an absolute magnitude of $M_{1,450}=-26.76\pm 0.04$. 
 To calculate the bolometric luminosity of the quasar ($L_{\rm Bol}$), we first fit a power-law continuum to the spectrum and measure the luminosity at rest-frame $3,000\,$\AA\ ($L_{3,000}$). We then use the bolometric correction\cite{richards2006b} $L_{\rm Bol} = 5.15 \times L_{3,000}$, which results  in $L_{\rm Bol}=4.0 \times 10^{13}\,L_\odot$, where $L_\odot$ is the luminosity of the Sun. 
 The \mgii\ line has a full-width at half maximum of $2,500^{+480}_{-320}\, \kms$, which together with the luminosity yields a black hole mass of $7.8^{+3.3}_{-1.9} \times 10^{8}\, M_\odot$, where $M_\odot$ is the mass of the Sun. The reported errors do not include the dominant systematic uncertainties in the local scaling relations\cite{vestergaard2009} of 0.55 dex. The accretion rate of this quasar is consistent with Eddington accretion, with an Eddington ratio of $L_{\rm Bol}/ L_{\rm Edd} = 1.5^{+0.5}_{-0.4}$. 
 
\begin{figure}[h!]
\centering
\includegraphics[width=\linewidth]{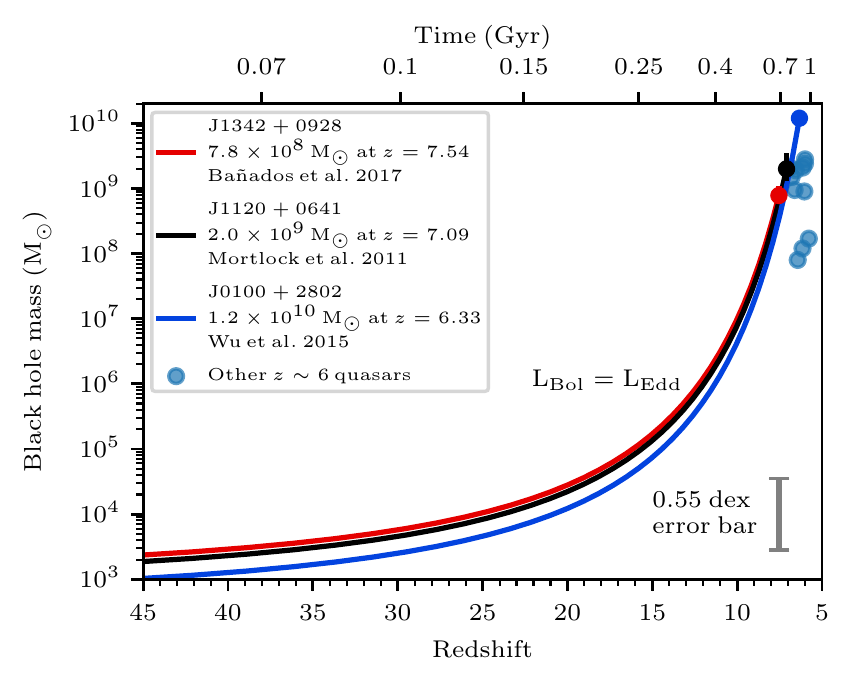}
\caption{\small
\textbf{Black-hole growth of three of the highest-redshift and most-massive quasars in the early Universe}. J1342+0928 at $z=7.54$ has a black-hole mass of $7.8\times 10^8\,M_\odot$, J1120+0641 at $z=7.09$ has a black-hole mass of $2.0\times 10^9\,M_\odot$ (ref. 1) , and J0100+2802  at $z=6.33$ has a black-hole mass of $1.2\times 10^{10}\,M_\odot$ (ref. 16). The three curves are normalized to the observed black-hole mass and redshift of these quasars (data points with statistical error bars). The black-hole growth is modelled as $M_{\rm BH}= M_{\rm BH,seed} \times \exp({\rm time}/ 50 {\rm \, Myr})$, where $t$ is time and we have assumed that the black holes are accreting at the Eddington limit ($L_{\rm Bol} = L_{\rm Edd}$) with a radiative efficiency of 10\%. The pale blue circles show a compilation\cite{gallerani2017} of black-hole masses of quasars at $z\approx 6$.  
The grey error bar at the bottom right represents dominant uncertainty, which is due to systematics in the local scaling relation that is used to estimate the black-hole mass of quasars at these redshifts\cite{vestergaard2009}. Ignoring this systematic uncertainty and assuming that the local relations apply to these extremely distant and luminous quasars, black-hole mass seeds more massive than  $1,000\, M_\odot$ by $z=40$ are necessary to grow the observed supermassive black holes in all three cases.
}
\label{fig:bhgrowth}
\end{figure}

 The existence of supermassive black holes in the early Universe poses crucial questions about their formation and growth processes. Observationally, the most distant quasars provide joint constraints on the mass of black-hole `seeds' and their accretion efficiency.  Assuming a typical matter-energy conversion efficiency\cite{pacucci2015} of 10\%, a black hole accreting at the Eddington rate grows exponentially on timescales of around 50\,Myr.  In Fig. \ref{fig:bhgrowth} we show the black-hole growth of three quasars assuming that they accrete at the Eddington limit throughout their entire life:  J$1342+0928$ at $z=7.54$, J1120+0641 at $z=7.09$ \cite{mortlock2011}, and SDSS~J0100+2802 at $z=6.33$\cite{wu2015}. 
 These are the quasars that currently place the strongest constraints on early black-hole growth.   
 In all three cases, black-hole seeds of at least $1000$\,$M_\odot$ are required by $z=40$. The existence of these supermassive black holes at $z>7$ is at odds with models of early black-hole formation that do not involve either massive (more than about $10^4\,M_\odot$) seeds or episodes of hyper-Eddington accretion.

The epoch of reionization was the last major phase transition in the Universe, when it changed from being completely neutral to ionized.  The presence of complete Gunn-Peterson troughs\cite{beckerG2015} in the spectra of $z\gtrsim 6$ quasars indicates that there were only traces of neutral hydrogen (volume-averaged fraction of neutral hydrogen $\fhi > 10^{-4}$) at $z\approx 6$. Because the Ly$\alpha$ transition saturates at larger neutral fraction, Gunn-Peterson troughs are sensitive to only the end phases of reionization\cite{beckerG2015}. Therefore, to probe the epoch when reionization occurred,  we need alternative methods. During earlier stages of reionization ($\fhi>0.1$), neutral intergalactic matter should produce the characteristic damped Gunn-Peterson absorption redwards of the Ly$\alpha$ emission line\cite{miralda-escude1998} (this region of the spectrum is hereafter referred to as the damping wing). Evidence of this long-sought signature in quasar spectra has been reported only once, in the previous redshift record holder at $z=7.09$\cite{mortlock2011,greig2016,bosman2015}. 

\begin{figure}[h!]
\centering
\includegraphics[]{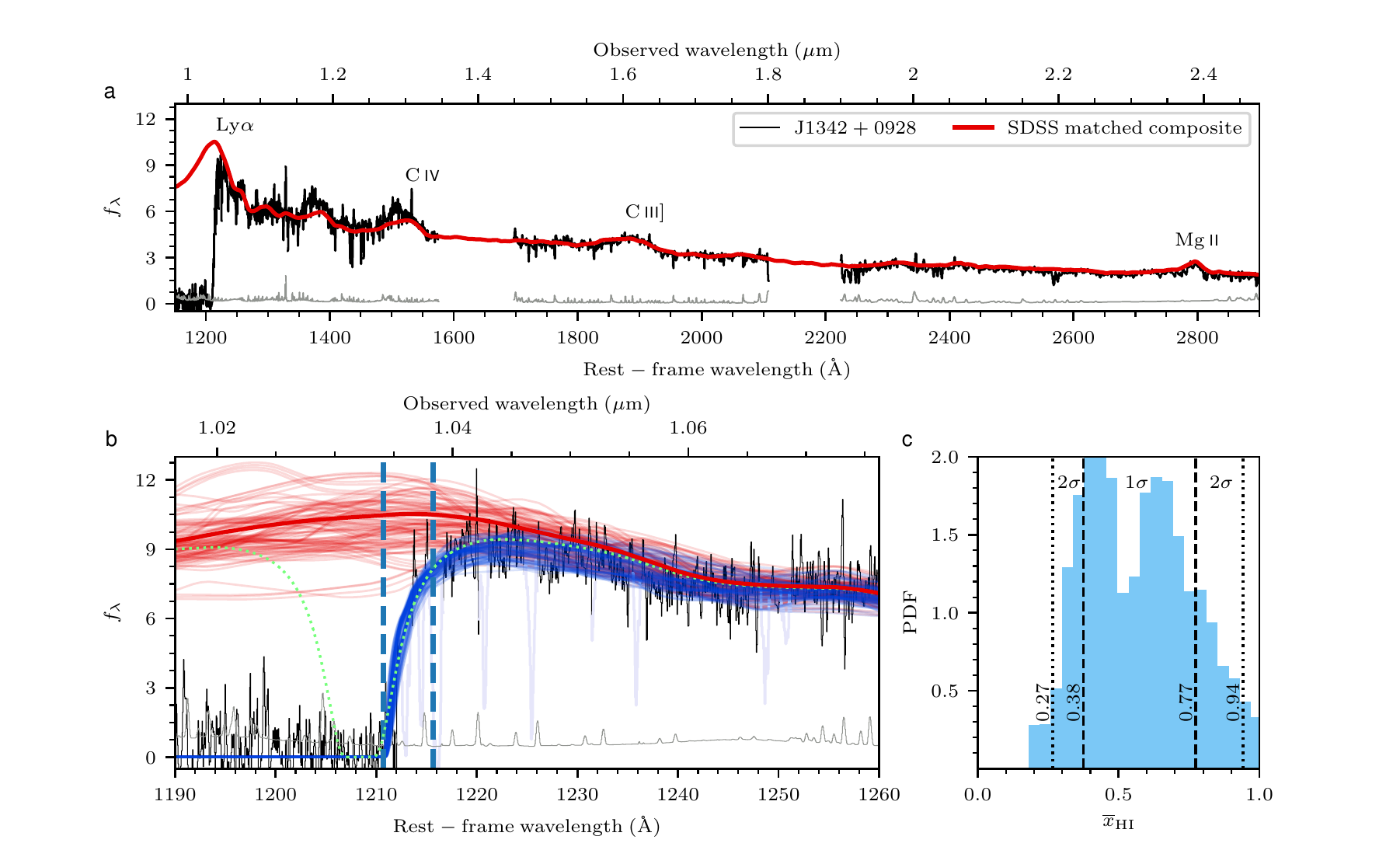}
\vspace{-1cm}
\caption{ \small
\textbf{Continuum emission and damping-wing modelling in the spectrum of J1342+0928} (in units of $10^{-18}\,\mathrm{erg}\; \mathrm{s}^{-1}\;\mathrm{cm}^{-2}\;$\AA$^{-1}$). 
\textbf{a,} The quasar spectrum (black and grey) is the same at that shown in Fig.~\ref{fig:pisco}. The continuum model (red) is constructed by averaging the SDSS DR12 quasars\cite{paris2017} that were not flagged as broad-absorption line quasars in the redshift interval $2.1<z_{\rm \mgii}<2.4$, and that have a median signal-to-noise ratio greater than $5$ in the \civ\ region. The subsample was further refined by considering only SDSS quasars with a \civ\ blueshift with respect to \mgii\ within $1,000\,\kms$ of that observed in J$1342+0928$ ($6,090 \pm 275\, \kms$) and with rest-frame \civ\ equivalent widths consistent at the $3\sigma$ level with the \civ\ equivalent width of J$1342+0928$ ($11.3 \pm 0.8\,$\AA). This yielded 46 `analogue' quasars.  Their continua were fitted individually by a slowly varying spline to remove strong absorption systems and noisy regions, and to interpolate between high-transmission peaks bluewards of Ly$\alpha$. We then normalized each spectrum  at $1,290\,$\AA\ and averaged them. This composite spectrum is shown in red, and reproduces fairly well the spectral features of J$1342+0928$. 
Assuming the systemic redshift derived from \cii, the proximity zone of J$1342+0928$, which is defined\cite{fan2006b,eilers2017} as the physical radius at which the transmission drops to 10\%, is 1.3 Mpc.
\textbf{b,} 
Close-up of the Ly$\alpha$ region, showing a strong absorption profile that can be modelled as a damping wing caused by a significantly neutral IGM. The black and grey curves are the FIRE spectrum and $1\sigma$ error, binned by a factor of two. The pale blue lines show regions masked out due to intervening absorption systems. The thick red line is the  SDSS-matched composite spectrum (see \textbf{a}); the thin red lines are used as models of intrinsic emission (100 out of 3,000 shown). To take into account the error in the prediction for any given quasar, these models consist of bootstrapped mean composite spectra plus a relative error vector (SDSS quasar - mean)/mean chosen randomly from the 46 possible error vectors.   The blue lines represent the expected\cite{miralda-escude1998} damping wing  assuming a fully ionized proximity zone (region between the vertical dashed lines), a constant neutral fraction $\fhi$ between the end of the proximity zone and $z=7.0$, and a fully ionized IGM at $z<7$. The exact choice of the transition redshift does not  greatly affect the results.  
The green dotted line is the absorption that would be caused by a single absorber with  $N_{\mathrm{HI}}=10^{20.5}\,$cm$^{-2}$ at $z = 7.49$.  
\textbf{c,} The derived  probability density function (PDF) for $\fhi$, with a median value and $68\%$ central interval of $\fhi = 0.56^{+0.21}_{-0.18}$.
In Methods we present one additional model of the intrinsic emission from the quasar and two additional models of the damping wing. All of our analyses require a significantly neutral IGM to reproduce the damping-wing profile in the spectrum of J$1342+0928$ (see Extended Data Table 2).
}
\label{fig:wing}
\end{figure}

To calculate the implied fraction of neutral hydrogen in the IGM ($\fhi$), we must first estimate the shape of the unabsorbed continuum and then fit a parameterized absorption model using the data and continuum as inputs.  This analysis is challenging because assumptions about the process of reionization need to be made and estimating the intrinsic strength of the Ly$\alpha$ emission for one single quasar is not straightforward. The latter is particularly difficult for the case of J$1342+0928$, which has extreme line blueshifts, which greatly reduces the number of lower-redshift quasars with which this source can be compared. We summarize our approach in Fig. \ref{fig:wing}: we  follow previous work\cite{mortlock2011,bosman2015} to estimate the intrinsic continuum by searching for lower-redshift quasars with similar spectral features to J$1342+0928$ and  a previously described method\cite{miralda-escude1998} to obtain an estimate of the neutral fraction. The main result is that a significantly neutral IGM is necessary to reproduce the damping-wing profile of J$1342+0928$. We find $\fhi = 0.55^{+0.21}_{-0.18}$, with the $95\%$ central interval of the $\fhi$ distribution of $0.26-0.93$ (see Fig. \ref{fig:wing}). 
To explore how robust this result is, we consider an alternative model for the intrinsic emission from the quasar and two more elaborate models of the  damping wing (Methods). All our analyses strongly favour a scenario in which the IGM surrounding J$1342+0928$ is significantly neutral, although the exact fraction of neutral hydrogen depends on the method (see Extended Data Table 2). Nevertheless, even our most conservative method indicates $\fhi > 0.11$  at 95\% probability. Higher signal-to-noise ratio and larger wavelength coverage of the spectrum of the quasar is necessary to refine and strengthen this result.

 An important caveat of our finding is that a similar absorption profile could also be caused by a single high-column-density absorber ($N_{\mathrm{\hi}} > 2\times 10^{20}\,$cm$^{-2}$) in the immediate vicinity of the quasar\cite{miralda-escude1998}. 
 Although we find a large number of foreground heavy-element 
absorbers at lower redshifts, we find no evidence for metal-line absorption at redshifts  near that of the quasar. Adopting the methodology of Simcoe et al. (2012)\cite{simcoe2012}, a single Ly$\alpha$ absorber at $z=7.49 \pm 0.01$ and $N_{\mathrm{\hi}} = 10^{20.50^{+0.32}_{-0.45}}\,$cm$^{-2}$ could produce the damping wing observed in J$1342+0928$. However, this absorber could have 
a metal abundance of at most $1/1,700$ times the solar value for oxygen ($95\%$ confidence), which would make it among the most distant and metal-poor absorbers known\cite{simcoe2012}. The probability of intercepting a discrete absorber with $N_{\mathrm{HI}} \gtrsim 10^{20.50}\,$cm$^{-2}$ within $2000\,\kms$ of a quasar at $z=7.5$ is less than $1\%$, based on the number density of such systems at lower redshifts\cite{songaila2010}. This low probability 
supports the hypothesis that the absorption profile in J$1342+0928$ is instead probing the 
neutral IGM gas in the epoch of reionization.

\begin{figure}[h!]
\centering
\includegraphics[width=\linewidth]{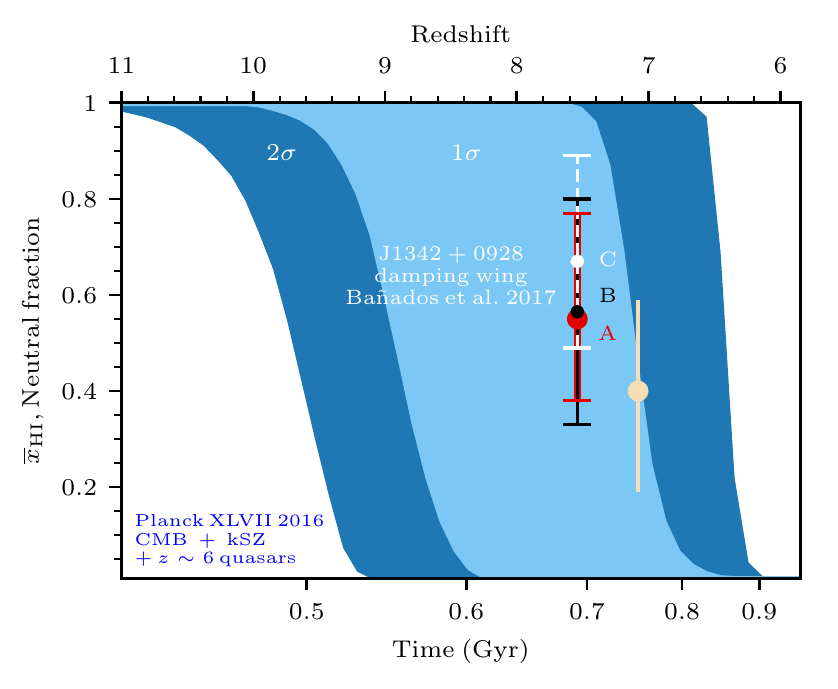}
\caption{\small 
\textbf{Constraints on the history of reionization}. The constraints are derived in terms of the fraction of neutral hydrogen ($\fhi$) and cosmic time from the Big Bang in gigayears (redshift $z$ is shown on the top axis). The contours show the 1$\sigma$ and $2\sigma$ constraints from  the optical depth\cite{planck2016XLVII} of the cosmic microwave background, the kinetic Sunyaev-Zeldovich (kSZ) effect, and quasars at $z\approx 6$ see figure 17 in ref. 12 and references therein).  The data points represent the median and  $1\sigma$ constraints from the damping-wing analyses of ULAS~J1120+0641 at $z=7.09$ (beige; $\fhi=0.40^{+0.21}_{-0.19}$; ref. 19) and J$1342+0928$ at $z=7.54$ (this work). The red point and error bar represents the analysis presented in Fig. \ref{fig:wing} (model A); the black and white measurements are the most conservative constraints (that is, lower neutral fraction) from the two additional models (B and C) of the damping wing of J$1342+0928$ presented in Methods (see Extended Data Table 2). All of our analyses consistently find a large fraction of neutral hydrogen surrounding J$1342+0928$. 
 The uncertainties in the damping-wing analyses do not include cosmic variance; that is, we are constraining one line-of-sight per quasar. A better understanding of the global history of reionization requires additional measurements at similar redshifts along more lines of sight. We note that the damping-wing constraints shown were obtained using a range of methods to perform the analysis and so have different systematics.
}
\label{fig:hi}
\end{figure}

Finally, the fact that both known quasars at $z>7$ show evidence of damping wings  confirms that we are starting to probe well within the epoch of reionization (see Fig. \ref{fig:hi}), in agreement with recent indications based on the number density of Ly$\alpha$-emitting galaxies at similar redshifts\cite{zheng2017} and results from the cosmic microwave background\cite{planck2016XLVII}.

\bibliography{library_nameyyyy}

\medskip
\medskip

\noindent \textbf{Acknowledgments:} 
We thank D. Ossip for his support with the FIRE echellete observations and A. Stephens for help preparing the GNIRS observations. 
This work is based on data collected with the Magellan Baade telescope, the Gemini North telescope (program  GN-2017A-DD-4), the Large Binocular Telescope, and the IRAM/NOEMA interferometer. We are grateful for the support provided by the staff of these observatories. We acknowledge the use of the UKIDSS, \textit{WISE}, and DECaLS surveys.  

\medskip
\noindent \textbf{Author contributions:} 
E.B., R.D., X.F., E.P.F., C.M., H.-W.R., D.S., B.P.V., F.Walter, F.Wang, and J.Y. discussed and planned the candidate selection and observing strategy, and analysed the data. E.B. selected the quasar and with D.S. obtained and analysed the discovery spectrum. R.A.S. provided the final FIRE data reduction. J.F.H. provided the final GNIRS data reduction. G.C.R. carried out the follow-up Fourstar observations for this quasar. D.D.K. reduced the follow-up Fourstar data.  J.M.W., B.P.V., and F.Walter contributed with the observations and analysis of the IRAM/NOEMA data. The damping wing analyses were carried out by E.B. (model A), F.B.D. and J.F.H (model B) and R.A.S. and M.L.T. (model C). F.B.D. and J.F.H. performed the PCA continuum modelling.  R.A.S. performed the analysis to find the characteristics of a single absorber that could cause the absorption profile of the quasar.  E.B. led the team and prepared the manuscript.  All co-authors discussed the results and provided input to the paper and data analysis.  

\medskip
\noindent \textbf{Author information:} 
Reprints and permissions information is available at www.nature.com/reprints. The authors declare no competing financial interests. Correspondence and requests for materials should be addressed to ebanados@carnegiescience.edu.

\newpage

\newpage

\newpage

\newpage 

\clearpage
%\newpage
\section*{Methods}

\subsection*{Damping-wing modelling}

To calculate the neutral fraction $\fhi$ of the IGM, we must first estimate the intrinsic emission from the quasar. After that, a model is fitted that reproduces the observed damping wing in the Ly$\alpha$ region.  We have shown that the data strongly suggest that the IGM  surrounding the quasar J$1342+0928$ at $z = 7.54$ is significantly neutral: $\fhi = 0.56^{+0.21}_{-0.18}$ (see Fig. 3). Recovering the intrinsic continuum of J$1342+0928$ is particularly challenging because of its extreme emission-line blueshifts, which are not well represented in matched composites at lower redshifts. Specifically, in the analysis presented here we found only 46 lower-redshift quasars with similar \civ\ blueshifts and equivalent widths. Furthermore, we have modelled the damping wing following a prescription\cite{miralda-escude1998} that assumes a homogeneous neutral IGM between the  proximity zone of the quasar and an arbitrary redshift, which we set to $z_{\rm END} = 7.0$,  and that considers the Universe to be completely ionized below that redshift.  We remark that given that the damping wing is produced by the IGM in the environment of the quasar, the results are insensitive to the exact value of $z_{\rm END}$. The results change by less than 1\% even in the extreme case of $z_{\rm END}=6.0$. Hereafter, we refer to the IGM model presented in the main text as model A.  

Here we reanalyse the data and present an additional model of the intrinsic emission from the quasar using a principal component analysis (PCA) decomposition. Furthermore, we introduce two completely independent methods to model the damping wing using the intrinsic emission from the quasar obtained in the main text (SDSS matched composite, hereafter continuum 1) and from the PCA model (hereafter continuum 2; described in more detail below) as inputs. In this way, we test how sensitive our conclusions are to different assumptions and systematics in the analysis. 

\subsection*{PCA model}

In this model, we predict the intrinsic quasar continuum in the Ly$\alpha$ region ($\lambda_{\mathrm{rest}}  < 1,260$\,\AA) on the basis of the remainder of the spectrum using a PCA analysis trained on $12,764$ quasar spectra from the SDSS/BOSS DR12 quasar catalogue\cite{paris2017}. Quasars in the training set were chosen to have signal-to-noise ratios greater than $7$ at $\lambda_{\mathrm{rest}}  < 1,285$\,\AA\ and to lie at redshifts $2.1 < z < 2.5$ such that the BOSS spectral range comfortably covered the Ly$\alpha$ region (down to $\lambda_{\mathrm{rest}} =1,190$\,\AA) and the \mgii\ region up to $\lambda_{\mathrm{rest}}  \approx 2,800$\,\AA, similar to the rest-frame spectral coverage of J$1342+0928$. We constructed red-side and blue-side PCA basis spectra from the training set after applying automated spline fits to recover the unabsorbed continuum. For each quasar in the training set, we fitted for the PCA coefficients and a template redshift simultaneously, placing each quasar into a common ``PCA redshift frame'' independent of (but largely similar to) their catalogue redshifts. We used these fits to define the matrix projection from red-side to blue-side PCA coefficients as in Suzuki et al. (2005)\cite{suzuki2005} and Pâris et al. (2011)\cite{paris2011}. We then fitted the red-side PCA coefficients of J1342+0928 in the same way as the training set. To estimate the bias and uncertainty of the continuum fit for quasars similar to J1342+0928, we measured the relative continuum error for the 1\% of quasars in the training set with the most similar red-side PCA coefficients. The relative uncertainty in the Ly$\alpha$ region for this subset was found to be about $7\%$. 

In Fig. E1 we compare the intrinsic Ly$\alpha$ emission from the quasar reconstructed by both the PCA and SDSS-matching analyses.  Continuum 2 predicts a slightly stronger emission in the Ly$\alpha$ region, but in both cases the emission is much weaker than for an average low-redshift SDSS quasar. 

\subsection*{Damping wing}

Using continuum 2 (PCA) as input for model A, we require an almost completely neutral Universe ($\fhi \approx 1$) to model the damping wing in the spectrum of J$1342+0928$. This is driven by the higher intrinsic flux in the Ly$\alpha$ region of continuum 2. We would obtain an even more dramatic result if we naively used an average SDSS quasar as input, which would have a much stronger Ly$\alpha$ emission line (Fig. E1). These results seem to reinforce our finding that the Universe is significantly neutral at $z\approx 7.5$. To assess whether this result depends on the method used to reproduce the damping wing, we now  introduce two more elaborate methods (hereafter models B and C).

\subsubsection*{Model B}

 In this model, we place the quasar within simulated massive dark-matter haloes at $z = 7.5$, which in turn populate large-scale over-densities that are already ionized by galaxies before the quasar turns on\cite{alvarez2007}. 
The damping-wing strength is then sensitive to the distance to the nearest neutral patch set by the morphology of reionization and to the output of ionizing photons from J$1342+0928$. Small proximity zones can result either from a high $\fhi$ in the surrounding IGM or from a short active lifetime of the central quasar\cite{eilers2017}. We modelled the residual \hi\ absorption inside the proximity zone of the quasar and the damping-wing profile from the neutral IGM by simulating 
the radiative transfer of ionizing photons through the IGM\cite{davies2016a} 
from the locations of massive dark-matter haloes using a realistic distribution of densities from a large-volume hydrodynamical simulation\cite{lukic2015}. The inside-out morphology of reionization as a function of $\fhi$ was computed from independent large-volume semi-numerical simulations of patchy reionization using a modified version of the 21cmFAST code\cite{mesinger2011}. 

The neutral-hydrogen fraction $\fhi$ and quasar lifetime $t_Q$ were then jointly constrained via a Bayesian approach using pseudo-likelihood in the spirit of indirect inference methods (see, for example, Drovandi et al. 2015\cite{drovandi2015}).  We define our pseudo-likelihood as the product of independent flux PDFs evaluated in spectral bins of 500\,\kms, and treat the set of maximum pseudo-likelihood parameter values ($\fhi$ and $t_Q$) as a summary statistic. 

We applied model B to the quasar continuum obtained from the PCA method described above (continuum 2), using millions of forward-modelled mock spectra, including a self-consistent treatment of the highly covariant continuum (determined from our PCA training set), to compute the posterior PDF.  Marginalizing over quasar lifetimes with a log-uniform prior in the range $10^5 < t_Q < 10^8$ yr, the central 68\% (95\%) credible interval for $\fhi$ is $0.45 - 0.87$ ($0.22 - 0.98$).  We show a representation of model B using the intrinsic emission from continuum 2 and the associated marginalized posterior PDF of  $\fhi$ in Fig. E2.

If we apply model B using continuum 1 as input, then the central 68\% (95\%) credible interval for $\fhi$ is $0.33 - 0.80$ ($0.11 - 0.96$); we show the associated posterior PDF of  $\fhi$ in Fig. E3a.

\subsubsection*{Model C}

In this model, the IGM absorption profile is modelled on the basis of the methods outlined in Bolton et al. (2011)\cite{bolton2011}. After the quasar turns on, it evacuates an expanding ionized region (the proximity zone) within the surrounding IGM. Its absorption profile is then specified by four parameters: (1) the ionizing luminosity of the quasar (constrained by photometry),  (2) the size of the proximity zone (which is related to the age of the quasar), (3) the mean density of the surrounding medium, 
and (4) $\fhi$ outside the proximity zone. We use an affine-invariant Markov Chain Monte Carlo (MCMC) solver\cite{foreman-mackey2013} to fit these four parameters. The ionizing luminosity of the quasar is estimated using the power-law indices from Telfer et al. (2002)\cite{telfer2002} and the quasar magnitudes from this work. We impose a Gaussian prior with a width determined from the errors on the power-law index from ref. 37; the remaining parameters are all given flat priors.

Before fitting our model, we applied an automated clipping procedure to remove spectral absorption features. We divided the normalized spectrum into bins of size 2.5\,\AA\ in the rest frame, and interpolated a B-spline through the mean flux in each bin. Any pixels with flux values more than $3\sigma$ below or $7\sigma$ above the interpolated values were masked. This procedure was repeated until convergence was achieved.  Then, we ran two MCMC realizations, one for each mean quasar continuum created above (1 and 2; see Fig. E1).  For each continuum model, we ran 100 chains of $2,000$ steps each, and used the final 200 steps to construct the posteriors (the burn-in occurs within the first 250 steps or so). 

For continuum 1, the central 68\% (95\%) credible interval for $\fhi$ is $0.49 - 0.89$ ($0.31 - 0.99$); its posterior PDF is shown in Fig. E3b. By contrast, model C requires a neutral Universe ($\fhi \sim 1$) if continuum 2 is used as input, in line with the result obtained using model A. 

\subsection*{Final remarks}

In Extended In Extended Data Table 2 we summarize the constraints on the neutral fraction obtained from all IGM modelling methods used here with the two different models of the  intrinsic emission from the quasar.  In all cases we recover a large neutral fraction in the IGM.  We show a comparison between these three models inf Fig.~4. To be conservative, we show the constraints obtained by using continuum 1 as input because it predicts weaker emission around the Ly$\alpha$ line than does continuum 2 (Fig. E1) and so lower neutral fractions are needed to explain the absorption profile in J$1342+0928$ (Extended Data Table 2).  The most conservative of the analysis is model B, which implies $\fhi  > 0.11$ at the $2\sigma$ level (see Fig. E3a)---one of strongest constraints yet in the history of reionization.

\medskip
\medskip
\medskip

\noindent \textbf{Data availability:} The datasets generated and analysed during this study are available from the corresponding author on reasonable request. \\
%\medskip

\noindent \textbf{Code availability:} We have opted not to make available the codes used to model the damping wing because they will be described in more detail and made available in forthcoming papers (model B, F.B.D. et al., manuscript in preparation; model C, M.L.T et al., manuscript in preparation).

\clearpage

%\newpage 

\renewcommand{\thefigure}{E\arabic{figure}}
\setcounter{figure}{0}

\begin{figure}[ht!]
\centering
\includegraphics[width=0.75\linewidth]{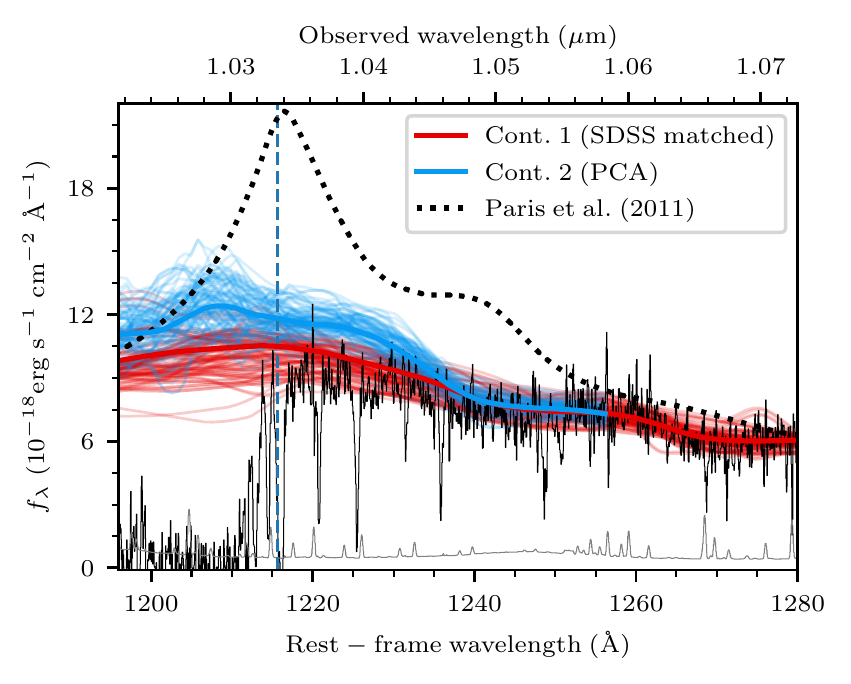}
\caption{\small 
\textbf{Modelling the intrinsic emission from J1342+0928.} The red lines represent the continuum used in the main text, which was constructed by averaging SDSS DR12 quasars with similar \civ\ properties (equivalent widths and blueshifts) to those observed in J$1342+0928$  (see Fig. 3). The blue lines are 100 random draws of PCA-reconstructed intrinsic emission, as described in Methods.  In both cases, the mean intrinsic spectrum is shown as a thick line. The vertical dashed line shows the Ly$\alpha$ wavelength. The PCA-reconstructed spectrum has a stronger emission around the Ly$\alpha$ line than does the SDSS-matched reconstructed emission. The dotted line is the mean SDSS quasar from ref.~29, which has a much stronger Ly$\alpha$ line than that of any of our continuum models of J$1342+0928$.
}
\end{figure}

\begin{figure}[ht!]
\centering
\includegraphics[]{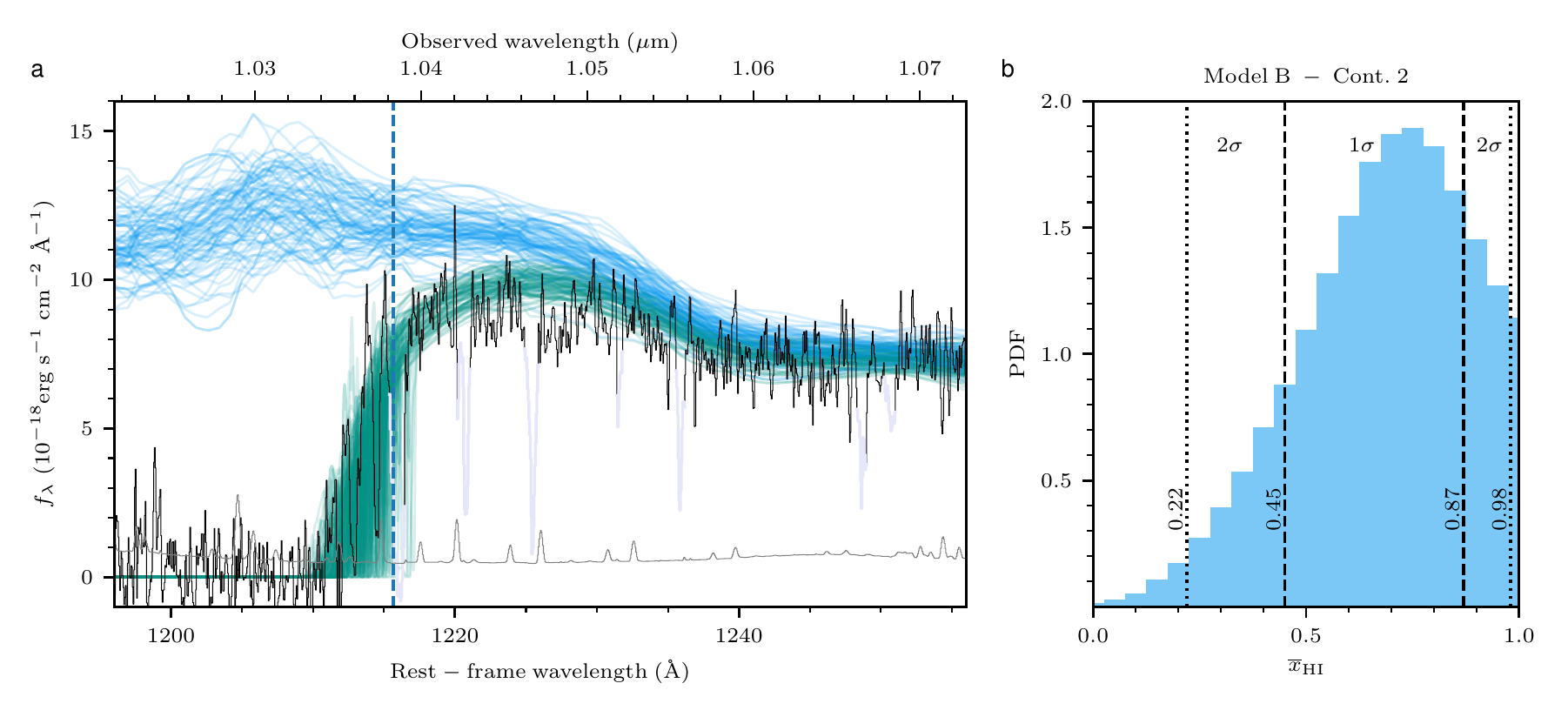}
\caption{\small 
\textbf{Damping-wing analysis with continuum 2 (PCA) and model B. a,} Same as Fig. 3b, but showing 100 realizations of the PCA-predicted intrinsic emission (blue) and damping-wing (green) draws from the posterior PDF of model B (see Methods for details). Model B masks absorption systems only redwards of the Ly$\alpha$ line (pale blue) because this model takes into account the internal absorption in the proximity zone, which explains the larger scatter bluewards of the Ly$\alpha$ line (dashed vertical line). \textbf{b,} The marginalized posterior PDF of $\fhi$. The 50th percentile is $\fhi =0.68$ and the 16th – 84th (2.5th – 97.5th) percentile interval is $0.45 - 0.87$ ($0.22 - 0.98$).
}
\end{figure}

\begin{figure}[ht!]
\centering
\includegraphics[]{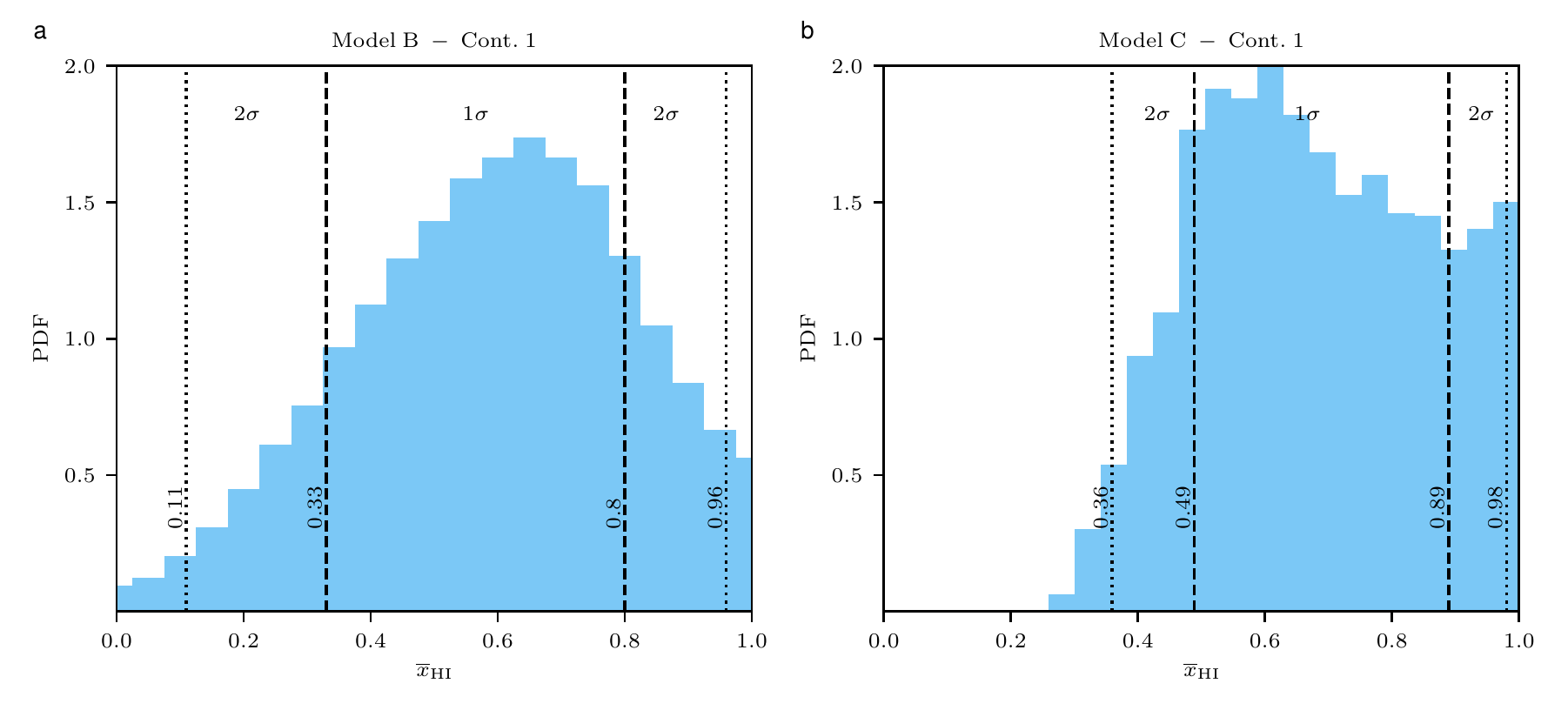}
\caption{\small \textbf{The marginalized posterior PDF of $\fhi$ using continuum 1 (SDSS-matched). a,} Model B. \textbf{b,} Model C.  Model B applied to continuum 1 yields the most conservative distribution of our analyses. Even in this case, a significantly neutral Universe with $\fhi > 0.11$ at the $2\sigma$ level is implied.
}
\end{figure}

%%%%%%%%%%%%%%%%%%%%%%%%%%

\renewcommand{\thetable}{E\arabic{table}} 
\setcounter{table}{0}

\begin{table}[h!]
\small
\centering
\caption{\textbf{Survey photometry of the quasar J1342+0928 at $\mathbf{z=7.54}$.}}
\begin{tabular}{lll}
\hline 
Survey  & AB  magnitudes&  \\
\hline
\vspace{2pt}

DECaLS & $z_{\rm DE, 3\sigma}>23.32$&            \\

\hline

\vspace{2pt}

UKIDSS &$Y=21.47\pm 0.19$ &$J=20.75\pm 0.11$     \\
	   &$H=20.02\pm 0.02$ &$K=20.03\pm 0.12$     \\
       \hline 

\vspace{2pt}
WISE   &$W1=20.17 \pm 0.15$ & $W2=20.11 \pm 0.29$          \\
\hline
\end{tabular}
\end{table}

\begin{table}[h!]
\small
\centering
\caption{\textbf{Summary of the constraints on the neutral fraction in the IGM surrounding  J1342+0928.}
 Constraints on $\fhi$ from the modelling of the damping wing of the quasar using three different IGM models and two different intrinsic-emission models. The central 68\% (95\%) credible intervals are reported, except when a completely neutral IGM ($\fhi \approx 1$) is always the preferred solution .
}
\begin{tabular}{llll}
\hline 
\backslashbox{Continuum}{IGM Model} & A -- 68\% (95\%)&  B -- 68\% (95\%)  &  C -- 68\% (95\%)\\
%& A   & B&  \\
\hline %\\[1pt]
1 (SDSS-matched) & $0.38-0.77$ ($0.27-0.94)$ &   $0.33-0.80$ ($0.11-0.96)$   &   $0.49-0.89$ ($0.36-0.98)$       \\
\hline %\\[1pt]

%\vspace{3pt}

2 (PCA) &$\sim 1$ & $0.45-0.87$ ($0.22-0.98)$ & $\sim 1$    \\

\hline
\end{tabular}
\end{table}

\end{document}